\documentclass[11pt]{article}
\usepackage{amsmath,amssymb}

\usepackage{amsfonts}
\usepackage{amssymb,color}

\makeatletter \@addtoreset{equation}{section} \makeatother

\begin{document}

\title {Dirac-K\"{a}hler particle in Riemann spherical space:
\\ boson interpretation}
\author{  A.M. Ishkhanyan, O. Florea,  E.M. Ovsiyuk,  V.M. Red'kov}
\maketitle

\begin{abstract}

In the context of the composite boson interpretation, we construct
the exact general solution of the Dirac--K\"{a}hler equation
for the case of the spherical Riemann space of constant positive curvature,
for which due to the geometry itself one may expect
to have a discrete energy spectrum.
In the case of the minimum value of the total angular momentum, $j=0$,
the radial equations are reduced to second-order ordinary differential
equations, which are straightforwardly solved in terms of the hypergeometric
functions. For non-zero values of the total angular momentum, however,
the radial equations are reduced to a pair of complicated fourth-order differential equations.
Employing the factorization approach, we derive the general solution
of these equations involving four independent fundamental solutions
written in terms of combinations of the hypergeometric functions.
The corresponding discrete energy spectrum is then determined
via termination of the involved hypergeometric series,
resulting in quasi-polynomial wave-functions.
The constructed solutions lead to notable observations
when compared with those for the ordinary Dirac particle.
The energy spectrum for the Dirac-K\"{a}hler particle
in spherical space is much more complicated. Its structure
substantially differs from that for the Dirac particle
since it consists of two paralleled energy level series
each of which is twofold degenerate. Besides, none of the
two separate series coincides with the series for the Dirac particle.
Thus, the Dirac--K\"{a}hler field cannot be interpreted
as a system of four Dirac fermions.
Additional arguments supporting this conclusion are discussed.

\vspace{5mm}



\end{abstract}

\section{Introduction}

The concept of elementary particles as certain relativistically invariant objects
in the frame of the 4-dimensional Minkowski space-time is well appreciated
already for a long time. For any particle, it assumes
certain transformation properties of a corresponding field,
and a wave equation that describes this field.

An interesting problem here, that attracted much attention in past,
is the extension of the wave equation for the Dirac--K\"{a}hler field.
The literature devoted to this field is huge (see, e.g., [1--33]).
The discussion started already with the early development
of the theory of quantum mechanical wave equations,
just after the concept of a particle with spin 1/2 appeared \cite{1928-Dirac}.
The Dirac--K\"{a}hler field was invented yet in 1928--1929 by Darwin \cite{Darwin},
Ivanenko and Landau \cite{1928-Ivanenko} as an alternative to the Dirac field.
The main intention was to construct a wave equation for a spin 1/2 particle
on the basis of tensor objects without using spinors.
In particular, an argument was that the new objects, the spinors,
seemed mysterious and obscure in comparison to the familiar tensors.
The Dirac--K\"{a}hler field consists of a set of tensors, which is equivalent to a 2-rank bispinor.
It contains 16 independent components, and its  wave equation can be presented formally
as four not-connected Dirac-type equations.
The main feature of the Dirac--K\"{a}hler field is that
it gives a possibility to perform (seemingly) smooth transition from tensors to spinors,
however, in a sense, it was an attempt to eliminate spinors at all.
This was the main ground to consider this object as related to fermions.
However, the mentioned disconnection of the four involved Dirac-type equations
is valid only for the case of flat Minkowski space-time,
and this feature is not preserved in the presence of a gravitational field,
that is for any non-Euclidean space-time model.

The three most interesting points in connection with the general covariant extension of the wave equation for this field are:
first, in flat Minkowski space there exist tensor and spinor formulations of the theory;
 second, in the initial tensor form there are tensors with different intrinsic parities;
   and third, there exist different views about physical interpretation of the object:
whether it is a composite boson or a set of four fermions.

In the Minkowski space, the Dirac--K\"{a}hler  particle is
described by 16-component wave function $U(x)$, a bispinor of rank 2 , or
by equivalent set of tensor fields: $ \{ \Phi (x), \Phi _{i}(x),
\tilde{\Phi }(x),
     \tilde{\Phi }_{i}(x), \Phi _{mn}(x)  \} $, where
      $\Phi (x)$ is a scalar, $\Phi _{i}(x)$ is a vector,
$\tilde{\Phi }(x)$ represents a \index{pseudo-scalar}
pseudo-scalar, $\tilde{\Phi }_{i}(x)$ represents a
\index{pseudo-vector} pseudo-vector, $\Phi _{mn}(x)$ is an
antisymmetric tensor. Connection between these quantities  is given by the
formula \cite{Ovsiyuk-Kisel-Redkov}
\begin{eqnarray}
U  = \left (  - i  \Phi   + \gamma^{l}  \Phi _{l}  +
           i  \sigma^{mn}   \Phi _{mn}  +  \gamma ^{5}  \tilde{\Phi }  +
           i \; \gamma ^{l} \gamma ^{5}  \tilde {\Phi }_{l}  \right  )  E^{-1} \; ,
\label{1.1}
\end{eqnarray}

\noindent where $ \gamma^{5} = - i \gamma^{0} \gamma^{1}
\gamma^{2} \gamma^{3} \; , \; \sigma^{ab} = {1 \over 4}
(\gamma^{a} \gamma^{b} - \gamma^{b} \gamma^{a}) \; $, and $E$
stands for bispinor metrical matrix \cite{Ovsiyuk-Kisel-Redkov}.

For a curved space-time background, the  generally covariant
tetrad-based Dirac--K\"{a}hler  equation in 4-spinor form
has the form \cite{Ovsiyuk-Kisel-Redkov}
\begin{eqnarray}
[\;  i \gamma ^{\alpha }(x) \;  ( \partial /\partial x^{\alpha } +
B_{\alpha }(x) ) - m \; ]\; U(x) = 0 \; , \label{1.2}
\end{eqnarray}

\noindent where  $B_{\alpha }$ is the  2-bispinor connection:
$ B_{\alpha }  = {1\over 2} J^{ab} e^{\beta }_{(a)}
 \nabla _{\alpha }(e_{(b)\beta }) $
with   $J^{ab} =  \sigma ^{ab} \otimes  I
+ I \otimes \sigma ^{ab} $ standing for generators of second rank
bispinor under the Lorentz group.
This spinor  equation is equivalent to the following generally covariant
tensor system \cite{Ovsiyuk-Kisel-Redkov}:
\begin{eqnarray}
\nabla ^{\alpha } \Psi _{\alpha }    + m \Psi  = 0 \; , \qquad
\nabla ^{\alpha } \tilde{\Psi} _{\alpha} + m \tilde{\Psi}   = 0 \;
, \nonumber
\\
\nabla _{\alpha }\Psi  + \nabla ^{\beta } \Psi _{\alpha \beta } -
 m \Psi _{\alpha }= 0 \; ,
\nonumber
\\
 \nabla _{\alpha } \tilde{\Psi}   -
{1\over 2} \epsilon ^{\;\;\beta \rho \sigma }_{\alpha }(x)
 \nabla _{\beta } \Psi _{\rho \sigma } -
m \tilde{\Psi }_{\alpha }  = 0 \; , \nonumber
\\
 \nabla _{\alpha } \Psi _{\beta } -
\nabla _{\beta } \Psi _{\alpha } + \epsilon ^{\;\;\;\;\rho \sigma
}_{\alpha \beta }(x) \nabla _{\rho }
 \tilde{\Psi }_{\sigma }  - m \Psi _{\alpha \beta } = 0 \; .
\label{1.4}
\end{eqnarray}

\noindent The covariant tensor field variables are connected with
the local tetrad tensor variables by the relations
\begin{eqnarray}
\Psi _{\alpha } = e^{(i)}_{\alpha } \Psi _{i} \; , \; \tilde{\Psi
}_{\alpha }= e^{(i)}_{\alpha } \tilde{\Psi }_{i} \; , \; \Psi
_{\alpha \beta } = e^{(m)}_{\alpha } e^{(n)}_{\beta } \Psi _{mn}
\; , \label{1.5}
\end{eqnarray}

\noindent and the Levi-Civita object is determined as
$
\epsilon ^{\alpha \beta \rho \sigma }(x) = \epsilon ^{abcd}
e^{\alpha }_{(a)}  e^{\beta }_{(b)}
                 e^{\rho }_{(c)}  e^{\sigma }_{(d)}   \; .
$
The fields  $\Psi  ,\; \Psi _{\alpha } , \; \Psi
_{\alpha \beta }$  are tetrad  scalars,   $\tilde{\Psi }, \;
\tilde{\Psi }_{\alpha }$  are tetrad  pseudo-scalars, the
Levi-Civita object $\epsilon ^{\alpha \beta \rho \sigma }(x)$  is
simultaneously  a generally covariant tensor and a tetrad
pseudo-scalar.

The most of the earlier work on this object concentrates on the symmetries
and other fundamental aspects of the relationships between
the Dirac and Dirac-K\"{a}hler fields [1--33], however, there are no known
non-trivial treatments for the Dirac--K\"{a}hler equation
when discussing the Dirac--K\"{a}hler field on curved space-time backgrounds.
However, the curved space-time models are of primary importance,
because on a non-Euclidean geometric background
only a composite boson interpretation for the Dirac--K\"{a}hler field is possible.

In the present paper we construct an exact solution for the case
of the simplest non-Euclidean geometrical model - the spherical
Riemann space of positive curvature. Due to the geometry itself,
one may expect to get discrete energy spectrum for a Dirac--K\"{a}hler particle.
The latter supposition turns out to be the case. Below we construct,
in terms of the hypergeometric functions, the general solution of the problem
consisting of four particular fundamental solutions,
each of which allows a separate discrete energy level series.
The determined spectrum for a Dirac--K\"{a}hler particle reveals substantial
peculiarities as compared with the spectrum for the ordinary Dirac case.
It exhibits a double-series structure, each of the series being twofold degenerate, and, moreover,
none of the energy level series coincides with that one for the Dirac particle.

\section{Separation of the variables}

The known results concerning the separation of the variables
for  the Riemann spherical space are as follows.
The Dirac-K\"{a}hler equation in the Minkowski space is written as \cite{Ovsiyuk-Kisel-Redkov}
\begin{eqnarray}
\left [ \; i \gamma ^{0}\; \partial _{t} \; + \;
 i \left ( \gamma ^{3} \; \partial _{r} \; +  \; { \gamma ^{1}J^{31} \; + \;
\gamma ^{2}\; J^{32} \over r} \right  ) \; + \; {1 \over r}\;
 \Sigma _{\theta ,\phi }\; - \; m\;  \right  ]\;  U(x) = 0 \; ,
\label{2.1.1a}
\\
\Sigma _{\theta ,\phi }  =  \; i \gamma ^{1}\; \partial _{\theta }
\; +  \; \gamma ^{2}\; {i\partial _{\phi }\; +\; iJ^{12} \; \cos
\theta \over  \sin  \theta} \;  , \qquad  J^{12} = ( \sigma ^{12}
\otimes  I \; + \;  I \otimes  \sigma ^{12} ) \; . \label{2.1.1b}
\end{eqnarray}

\noindent Diagonalizing the operators of the total angular momentum
\begin{eqnarray}
J_{1}  =  l_{1} \;+\; {i J^{12} \cos \phi \over  \sin \theta } \;
, \;\; J_{2} =   l_{2} \;+\; {i J^{12} \sin \phi \over  \sin
\theta }  \; , \;\; J_{3} = l_{3} \; , \label{2.1.2a}
\end{eqnarray}

\noindent for wave function we have the following general representation:
\begin{eqnarray}
U_{\epsilon jm}(t,r,\theta ,\phi ) = { e^{-i\epsilon t} \over  r}
\; \left | \begin{array}{llll}
f_{11} \; D_{-1} & f_{12} \; D_{0} & f_{13} \; D_{-1} & f_{14} \; D_{0} \\
f_{21} \; D_{0}  & f_{22} \; D_{+1}& f_{23} \; D_{0}  & f_{24} \; D_{+1} \\
f_{31} \; D_{-1} & f_{32} \; D_{0} & f_{33} \; D_{-1} & f_{34} \; D_{0} \\
f_{41} \; D_{0}  & f_{42} \; D_{+1}& f_{43} \; D_{0}  & f_{44} \;
D_{+1}
\end{array} \right |     ,
\label{2.1.2b}
\end{eqnarray}

\noindent where $f_{ab} = f_{ab}(r)$, $ D_{\sigma } =
D^{j}_{-m,\sigma } (\phi ,\theta ,0)$ stands for the Wigner
functions \cite{VMX}, and $j$ takes  the values $0, \; 1, \; 2, \ldots$  .
The next step is to diagonalize  the
space reflection operator. In the spherical tetrad basis it has
the form
\begin{eqnarray}
\hat{\Pi} =  \left | \begin{array}{cccc}
0 &  0 &  0 &  -1  \\    0 &  0 & -1 &   0 \\
0 & -1 &  0 &   0  \\   -1 &  0 &  0 &   0
\end{array} \right |     \otimes     \left | \begin{array}{cccc}
0 &  0 &  0 &  -1  \\    0 &  0 & -1 &   0 \\
0 & -1 &  0 &   0  \\   -1 &  0 &  0 &   0    \end{array} \right |
\otimes \hat{P} \; \; . \label{46}
\end{eqnarray}

\noindent The eigenvalue equation $ \hat{\Pi} U_{\epsilon jm} = \Pi \;
U_{\epsilon jm}$ imposes  the following restrictions:
\begin{eqnarray}
f_{31}  = \pm  f_{24} \;\;  , \;\; f_{32} = \pm  f_{23} \;\; ,
\;\; f_{33}  = \pm  f_{22} \;\;  , \;\; f_{34} = \pm  f_{21} \;\;
, \nonumber
\\
f_{41}  = \pm  f_{14} \;\; , \;\; f_{42} = \pm  f_{13} \;\; , \;\;
f_{43}  = \pm  f_{12} \;\; ,\;\; f_{44} = \pm  f_{11} \;\; ,
\label{47}
\end{eqnarray}

\noindent  where the upper sign refers to the eigenvalue $\Pi = (-1)^{j+1}$,
and the
lower sign refers to  $\Pi = (-1)^{j}$. Eq. (\ref{2.1.2b}) is rewritten as
\begin{eqnarray}
U_{\epsilon jm\delta } (t,r,\theta ,\phi ) = {e^{-i\epsilon t}
\over r} \left | \begin{array}{rrrr}
f_{11} \; D_{-1} & f_{12} \; D_{\;\;0}  & f_{13} \; D_{-1} & f_{14} \; D_{\;\;0}   \\
f_{21} \; D_{\;\;0}  & f_{22} \; D_{+1} & f _{23}\; D_{\;\;0}  & f_{24} \;  D_{+1} \\
\delta  f_{24} \; D_{-1}  & \delta  f_{23} \; D_{\;\;0}  &
\delta  f_{22} \; D_{-1}  & \delta  f_{21} \; D_{\;\;0}  \\
\delta  f_{14} \; D_{\;\;0}   & \delta  f_{13} \; D_{+1} & \delta
f_{12} \; D_{\;\;0}   & \delta  f_{11} \; D_{+1} \noindent
\end{array} \right |     \; , \label{48}
\end{eqnarray}

\noindent where $\delta=+1$ refers to the case $\Pi = (-1)^{j}$
and $\delta=-1$ stands for the case $\Pi = (-1)^{j+1}$.

The system of radial equations for $\delta = + 1$  reads
\begin{eqnarray}
\epsilon f_{24} - i {d\over dr} f_{24} +{i \over r} 0   - {ia
\over r}  f_{14}  - m f_{11} = 0 \; ,\quad \epsilon f_{23} - i
{d\over dr} f_{23} -{i \over r} f_{14}    - {ia \over r}  f_{13}
- m f_{12} = 0 \; , \nonumber
\\
\epsilon f_{14} + i {d\over dr}f_{14} +{i \over r} f_{23}    + {ia
\over r}  f_{24}  - m f_{21} = 0 \; , \quad \epsilon f_{13} + i
{d\over dr} f_{13} +{i \over r} 0    + {ia \over r}  f_{23}  - m
f_{22} = 0 \; , \nonumber
\\
\epsilon f_{22} - i  {d\over dr} f_{22} +{i \over r} 0    - {ia
\over r}  f_{12}  - m f_{13} = 0\; , \quad \epsilon f_{21} - i
{d\over dr} f_{21} -{i \over r} f_{12}    - {ia \over r}  f_{11}  -
m f_{14} = 0\; , \nonumber
\\
\epsilon f_{12} + i  {d\over dr} f_{12} +{i \over r} f_{21}    +
{ia \over r}  f_{22}  - m f_{23} = 0\; , \quad \epsilon f_{11} + i
{d\over dr} f_{11} +{i \over r} 0    + {ia \over r}  f_{21}  - m
f_{24} = 0 \; . \label{49}
\end{eqnarray}

\noindent The set of the equations for the case $\delta= -1$ is readily obtained from this system by the formal
change $m \to -m$.

Introducing the notations
\begin{eqnarray}
A = (f_{11} + f_{22})  \; , \quad iB = (f_{11} - f_{22})
 \;  ,\quad
 C = (f_{12} + f_{21})  \;, \quad  i D = (f_{12} - f_{21})
\; , \nonumber
\\
K = (f_{13} + f_{24})  \; , \qquad i L = (f_{13} - f_{24})
 \;  , \quad  M = (f_{14} + f_{23})   \; , \qquad i N = (f_{14} -
f_{23})  \; ,
 \label{50}
\end{eqnarray}

\noindent
the system (\ref{49}) is changed to a form without complex coefficients:
\begin{eqnarray}
\epsilon  K  - {dL \over dr} + {a \over r} N  - m A   = 0 \;
,\quad \epsilon  A - {dB \over dr} + {a \over r} D  - m K  = 0 \;
, \nonumber
\\
\epsilon  L  + {dK \over dr} + {a \over r} M  + m B  = 0 \; ,\quad
\epsilon  B + {dA \over dr} + {a \over r} C + m L  = 0  \; ,
\nonumber
\\
\epsilon  M - {dN \over dr} + {1\over r} N + {a \over r} L - m C =
0 \; ,\quad \epsilon  C - {dD \over dr} + {1 \over r} D + {a \over
r} B - m M = 0 \; , \nonumber
\\
\epsilon  N + {dM \over dr} + {1 \over r} M + {a \over r} K + m D
= 0 \; ,\quad \epsilon  D + {dC \over dr} + {1 \over r} C + {a
\over r} A + m N = 0 \; . \label{51}
\end{eqnarray}

These equations permit imposing of the following linear
constraints:
\begin{eqnarray}
A = \lambda \;  K \; , \;\; B = \lambda \; L \; , \;\; C = \lambda
\;  M \; , \;\; D = \lambda \;  N \;  \; \label{52}
\end{eqnarray}

\noindent with $\lambda  = \pm 1$. For the case  $\lambda = + 1$, instead of (\ref{51})
we then derive four equations:
\begin{eqnarray}
{dK \over dr} + {a \over r} M + (\epsilon  + m) L = 0 \; , \quad
{dL \over dr} - {a \over r} N - (\epsilon  - m) K = 0 \;  ,
\nonumber
\\
\left ({d \over dr} + {1 \over r} \right  ) M  + {a \over r} K + (\epsilon  + m)
N = 0 \; , \quad \left ({d \over dr} - {1 \over r} \right ) N  - {a \over r} L
- (\epsilon  - m) M = 0 \; . \label{53}
\end{eqnarray}

\noindent If here $m$ is changed to $-m$, one gets a corresponding set of four equations for
the case $\lambda = -1$.

Eqs. (\ref{53}) can be further reduced as follows:
\begin{eqnarray}
(i) \;\;  \sqrt{j+1} \; K(r) = f(r)\;\; , \; \sqrt{j+1} \;\; L(r) =
g(r) \; \; , \nonumber
\\
\sqrt{j}\; M(r) = f(r) \; , \; \sqrt{j} \; N(r) = g(r) \; ,
\nonumber
\\
 \left  ({d \over dr} + {j+1 \over r} \right ) f + (\epsilon + m) g  =
0 \; , \nonumber
\\
\qquad \;\; \left  ({d \over dr} - {j+1 \over r} \right ) g - (\epsilon - m) f =
0 \; ; \label{55}
\end{eqnarray}
\begin{eqnarray}
(ii) \;\;   \sqrt{j}\; K(r) = f(r)\;\; , \; \sqrt{j}\; L(r) = g(r)
\; , \nonumber
\\
\sqrt{j+1}\; M(r) = - f(r)\;  ,  \; \sqrt{j+1} \; N(r) = - g(r) \;
, \nonumber
\\
 \left ( {d \over dr}  - {j \over r} \right ) f + (\epsilon + m) g  =
0 \; , \nonumber
\\
\qquad \;\;\;
\left  ({d \over dr} + {j \over r}\right  ) g  -  (\epsilon - m) f
= 0 \;  . \label{56}
\end{eqnarray}

The derived equations apply to the case of nonzero $j$. The case $j=0$ needs a separate treatment because
the Wigner functions $D^{j}_{0, \pm 1}$  with $j=0$ are  not well posed.
The initial substitution in this case is simpler:
\begin{eqnarray}
U_{\epsilon 00}(t,r) = {e^{-i\epsilon t} \over r} \left |
\begin{array}{cccc}
0 & f_{12} & 0 & f_{14} \\ f_{21} & 0 & f_{23} & 0 \\
0 & f_{32} & 0 & f_{34} \\ f_{41} & 0 & f_{43} & 0
\end{array} \right |       \; .
\label{59}
\end{eqnarray}

\noindent Space reflection operator divides  the  solutions (\ref{59}) into two
classes:
\begin{eqnarray}
\Pi = + 1\; \;  ( \delta  = + 1  ), \quad f_{32} = + f_{23} \; ,\;\; f_{34} = + f_{21} \; , \;\; f_{41} = +
f_{14} \; , \;\; f_{43} = + f_{12}     \; ; \label{60}
\end{eqnarray}
\begin{eqnarray}
 \Pi = - 1 \;\; (\delta = -1),  \quad  f_{32} = - f_{23} \; ,
\;\; f_{34} = - f_{21} \; ,  \;\; f_{41} = - f_{14} \; , \;\;
f_{43} = - f_{12} \;  . \label{61}
\end{eqnarray}

\noindent Using the identity $\Sigma _{\theta ,\phi } \;
U_{\epsilon 00} =  0$, we get the radial system
\begin{eqnarray}
\epsilon\;  M  - {dN \over dr} + {N \over r} - m\; C  = 0 \; ,
\;\; \epsilon\;  N  + {dM \over dr} + {M \over r} + m\; D  = 0  \;
, \nonumber
\\
\epsilon\;  C  - {dD \over dr} + {D \over r} - m \; M  = 0  \; ,
\;\; \epsilon\;  D + {dC \over dr}  + {C \over r} - m \; N  =  0
\;   . \label{62}
\end{eqnarray}

\noindent This system is further reduced as follows:

\vspace{2mm}
 $ C= + M \; , \; D = + N \;  ( \lambda  = + 1 )
$
\begin{eqnarray}
 \left ( {d \over
dr} + {1 \over r}\right  ) M + ( \epsilon  + m ) N  = 0 \; , \qquad
\left  ( {d \over dr} - {1 \over r} \right ) N - ( \epsilon  - m ) M = 0  \;  ;
\label{63}
\end{eqnarray}

 $ C = - M , \; D = - N\;  (\lambda = - 1)  $
\begin{eqnarray}
\left ( {d\over dr}  + {1 \over r} \right  ) M + ( \epsilon  - m ) N = 0 \; ,
\qquad
\left ( {d \over dr} - {1 \over r} \right ) N - ( \epsilon  + m ) M = 0
\; . \label{64}
\end{eqnarray}

Thus, we have established the disconnection of the radial equations.
However, we stress that this is only possible in the Minkowski space.

The above development is readily   extended to the case of the spherical
Riemann space model. The metric is given as
\begin{eqnarray}
dS^{2}  =  dt^{2}  - d r ^{2} - \sin  ^{2} r  ( d\theta ^{2} +
\sin ^{2} \theta  d \phi ^{2} ) \;  , \label{2.20}
\end{eqnarray}

\noindent  and the  Dirac--K\"{a}hler  equation for this geometry
is written as
\begin{eqnarray}
\left [\; i \gamma ^{0} \; \partial_{t}\; + \; i \left  (\gamma
^{3}\;
\partial_{r} \;  + \;
 {\gamma ^{1} J^{31} + \gamma ^{2} J^{32} \over \tan r }       \right )\;
+ \;{1 \over \sin  r } \; \Sigma _{\theta ,\phi }  -  m  \right ]
\; U(x) =  0  \; . \label{2.21}
\end{eqnarray}
The quantum number of the total angular momentum adopts the
values $j = 0, 1,2, ...\;$ .

The calculations are much the same as above (see the
detailed derivation in  \cite{Ovsiyuk-Kisel-Redkov}), and
differences appear only in the explicit form of the final systems of radial equations.

For the case of minimum $j=0$, the problem is reduced to the system
\begin{eqnarray}
\epsilon\;  M  - {dN \over dr} + {N \over \tan r} - m\; C  = 0 \;
, \;\; \epsilon\;  N  + {dM \over dr} + {M \over \tan r} + m\; D =
0  \; , \nonumber
\\
\epsilon\;  C  - {dD \over dr} + {D \over \tan r} - m \; M  = 0 \;
, \;\; \epsilon\;  D + {dC \over dr}  + {C \over \tan r} - m \; N
=  0 \;   , \label{2.23}
\end{eqnarray}

\noindent while for greater total angular momenta, i.e. for $j=1,2,3,...$,
the problem reduces to the following more complicated equations
($a =\sqrt{j(j+1)}$):
 \begin{eqnarray} {dK \over dr}  + {a \over \sin r} M +
(\epsilon  + m) L  = 0 \; , \nonumber \\
 {dL \over d r} - {a \over
\sin r} N - ( \epsilon - m) K  = 0 \; , \quad \quad \nonumber
\\
\left ({d \over d r} + {1 \over \tan r} \right ) M + {a \over \sin
r} K + (\epsilon  + m) N  = 0 \; , \nonumber
\\
 \left ({d \over d r} - {1 \over \tan r } \right ) N  - {a \over \sin
r  } L  - (\epsilon  - m) M  = 0 \; . \label{2.22}
\end{eqnarray}

Note that the above analysis of the spherical model is also
applicable if an external spherically symmetric potential is added.
In this case the only needed formal change is
$\epsilon \; \to \; \epsilon - U(r)\; . $

\section{The case of minimum total angular momentum $j=0$ }

\noindent If $j=0$, the system (\ref{2.23}) is reduced using two substitutions:

\vspace{2mm}
 $ C= + M \; , \; D = + N \;  ( \lambda  = + 1 ),
$
\begin{eqnarray}
\left  ( {d \over
dr} + {1 \over \tan r} \right  ) M + ( \epsilon  + m ) N  = 0 \; ,\quad
\left  ( {d \over dr} - {1 \over  \tan r} \right  ) N - ( \epsilon  - m ) M = 0  \;  ,
\label{3.1}
\end{eqnarray}

 $ C = - M , \; D = - N\;  (\lambda = - 1)  $,
\begin{eqnarray}
\left ( {d\over dr}  + {1 \over  \tan r} \right ) M + ( \epsilon  - m ) N = 0
\; , \quad
 \left ( {d \over dr} - {1 \over \tan r}\right  ) N - ( \epsilon  + m ) M = 0
\; . \label{3.2}
\end{eqnarray}
The two first-order equations of system (\ref{3.1}) lead to the following second-order equations
 for $M$ and $N$:
\begin{eqnarray}
\left({d^{2}\over dr^{2}}+\epsilon^{2}-m^{2}-{1+\cos^{2}r\over
\sin^{2}r}\right)\,M=0\, , \label{3.3}
\end{eqnarray}
\begin{eqnarray}
\left({d^{2}\over dr^{2}}+\epsilon^{2}-m^{2}+1\right)\,N=0\,,
\qquad N = \mbox{const}\;e^{\pm i
\sqrt{\epsilon^{2}-m^{2}+1}\;r}\,. \label{3.4}
\end{eqnarray}

\noindent Eq.(\ref{3.3})
is solved in terms of the hypergeometric functions. Indeed,
passing to a new independent variable
 $x =
(1-\cos r)/2$:
\begin{eqnarray}
x(1-x){d^{2}M\over dx^{2}}+\left({1\over 2}-x\right)\,{dM\over
dx}+\left(\epsilon^{2}-m^{2}+1-{1\over 2x}-{1\over
2(1-x)}\right)M=0\,,  \label{3.6}
\end{eqnarray}

\noindent and applying  the substitution $M(x)=x^A(1-x)^B F(x)$
with $A=-{1\over 2}\,,\;1\,,\; B=-{1\over 2}\,,\;1\,, $
we arrive at the Gauss hypergeometric equation for  $F(x)=F(\alpha,\beta;\gamma;x)$ with parameters
\begin{eqnarray}
 \alpha=2 -
\sqrt{\epsilon^{2}-m^{2}+1}\,,\;  \beta=2
+\sqrt{\epsilon^{2}-m^{2}+1}\,,\; \gamma={5\over 2}\,. \label{3.9}
\end{eqnarray}
In order to have  a solution which is finite on the spherical space with $r \in [0,\pi]$,
one should take only the positive values $A=1$  and $B=1$.
The condition for the solutions to turn into polynomials,
 $\alpha = -n, \; n = 0,1,2, ...$, gives
the energy spectrum for the states with $j=0$:
\begin{eqnarray}
\epsilon^{2} - m^{2}  + 1 = (2 + n )^{2}  , \quad n =0,1,2,3, ... \, .
\label{3.10}
\end{eqnarray}

To accomplish the development,
one should determine the coefficient connecting the functions $M(r)$ and $N(r)$.
With straightforward calculations, we get the following result:
\begin{eqnarray}
N =  N_{0}\, \sqrt{x \,(1-x)} \,F(-n-1,\; 3+n; \;3/2;\; x)\,,\qquad \qquad
\nonumber\\
M = M_{0}\, x\, (1-x) \,F(-n,\; 4+n; \; 5/2; \;x), \quad M_{0}
/  N_{0}=- (2 / 3) \,(\epsilon-m)\,. \label{3.21}
\end{eqnarray}

\section{Analysis of the radial equations for $j=1,2, ...$
}

With the help of  the first two equations of the system  (\ref{2.22}) we eliminate
the functions $L$ and $N$:
\begin{eqnarray}
L=-\frac{1}{\epsilon +m}\left( \frac{d }{dx} K +\frac{a}{\sin x} M
\right) ,
\nonumber
\\
N=- \frac{1}{\epsilon +m}\left( \frac{d }{dx} M +\frac{1}{\tan x} M +\frac{a}{%
\sin x} K \right) ,
\end{eqnarray}

\noindent  and  arrive at   two  equations relating
$K$ and $M$  (we use the notation $\epsilon^{2} - m^{2} = p^{2},\;
p > 0 $):
\begin{eqnarray}
\left ( {d^{2} \over dx^{2}} +p^{2}  - {a^{2} \over \sin^{2} x}
\right )K
 = {2a \cos x \over \sin^{2} x} M\; ,
\nonumber \\
 \left ( {d^{2} \over dx^{2}}  + p^{2} +1  - {a^{2}
+2 \over \sin^{2} x} \right )M
 = {2a \cos x \over \sin^{2} x} K\;.
\label{4.17}
\end{eqnarray}

\noindent Rewriting Eqs.(\ref{4.17}) for the variable $x =
\cos^{2}r$, we obtain
\begin{eqnarray}
\left ( (1- x)  4x {d^{2} \over dx^{2} }  + 2(1 -  2x ){d \over
dx}
 +p^{2}  - {a^{2} \over 1 - x } \right ) K
 = {2a  \sqrt{x}  \over 1 - x } M \; ,\quad
\label{eq.1}
\end{eqnarray}
\begin{eqnarray}
\left ( (1- x ) 4x {d^{2} \over dx^{2} } +2 (1- 2x) {d \over dx}
  +p^{2} +1  - {a^{2} +2 \over 1 -x } \right ) M
 = {2a \sqrt{x}\over  1 - x } K \; .
\label{eq.2}
\end{eqnarray}

Near the point $x=1$, the asymptotic behavior of the solutions is given as

$$ M =M_{0} (1 - x )^{\gamma} \;, \;  K =K_{0} (1 - x )^{\gamma} \; .
$$
Eqs. (\ref{eq.1}) and (\ref{eq.2}) then lead to
\begin{eqnarray}
(4\gamma^{2} -2 \gamma -a^{2}) K_{0} -2a M_{0}=0 \; , \quad  -2a
K_{0} + (4\gamma^{2} -2 \gamma -a^{2} -2) M_{0} =0 \; .
\end{eqnarray}

\noindent Whence, we get four possible values for $\gamma$:
$
\gamma = j /  2, \; (j+2)/2 , \; (-j+1)/2 , \; (-j -1)/2 $.
 Only the positive values $\gamma = j/ 2,
\; (j+2)/2 $ may describe  bound states.

Elimination of $M(x)$ from Eqs.(\ref{eq.1})-(\ref{eq.2}) results in the
following fourth-order ordinary differential equation for $K(x)$:
\begin{eqnarray}
x^{2}{d^{4}K\over
dx^{4}}+\left (7x+5-{5\over1-x}\right )\,{d^{3}K\over
dx^{3}}+\left (10-{1\over 2}\,p^{2}+{p^{2}+a^{2}-28\over
2\,(1-x)}+{15-2a^{2}\over 4\,(1-x)^{2}}\right )\,{d^{2}K\over
dx^{2}}+ \nonumber
\\
+\left ({1\over 4\,x}+{3p^{2}-7\over
4\,(1-x)}-{3\,p^{2}+a^{2}-9\over 4\,(1-x)^{2}}+{a^{2}\over
4\,(1-x)^{3}}\right )\,{dK\over dx}+ \nonumber
\\
+\left ({p^{2}-a^{2}\over 8\,x}+{p^{2}-a^{2}\over
8\,(1-x)}+{p^{4}+2\,p^{2}-2\,a^{2}\over
16\,(1-x)^{2}}-{a^{2}(p^{2}-1)\over
8\,(1-x)^{3}}+{a^{2}(a^{2}-2)\over 16\,(1-x)^{4}}\right)\,K=0\,,
\nonumber
\\
\label{4.26}
\end{eqnarray}
In the similar manner, elimination of $K(x)$ gives the
following equation for $M(x)$:
\begin{eqnarray}
x^{2}{d^{4}M\over
dx^{4}}+\left ( 7x+5-{5\over1-x}\right ) \,{d^{3}M\over
dx^{3}}+\left (10-{1\over 2}\,p^{2}+{p^{2}+a^{2}-28\over
2\,(1-x)}+{15-2a^{2}\over 4\,(1-x)^{2}}\right )\,{d^{2}M\over
dx^{2}}+ \nonumber
\\
+\left ({1\over 4\,x}+{3p^{2}-6\over
4\,(1-x)}-{3\,p^{2}+a^{2}-9\over 4\,(1-x)^{2}}+{a^{2}\over
4\,(1-x)^{3}}\right )\,{dM\over dx}+ \nonumber
\\
+\left ( {p^{2}-a^{2}-1\over 8\,x}+{p^{2}-a^{2}-1\over
8\,(1-x)}+{p^{4}+2\,p^{2}-2\,a^{2}-3\over
16\,(1-x)^{2}}-{a^{2}(p^{2}-1)\over
8\,(1-x)^{3}}+{a^{2}(a^{2}-2)\over 16\,(1-x)^{4}}\right )\,M=0\,.
\nonumber
\\
\label{4.29}
\end{eqnarray}

\section{General solution via factorization}

\noindent A tool for studying such complicated equations
is the operator factorization method when one tries to present
the higher-order differential operators as products of lower-order ones.
In our case this leads to the general solution of the problem if one tries
the following factorization:
\begin{eqnarray}
\hat{K}_{4} K(x) = \hat{A}_{2} \hat{K}_{2} K(x) = \hat{A}_{2} f(x)
= 0 \;,
\nonumber
\\
\hat{M}_{4} M(x) = \hat{B}_{2} \hat{M}_{2} M(x) =  \hat{B}_{2}
g(x) = 0 \; .
\end{eqnarray}

\noindent Obviously, the solutions of the second-order equations
$$
f(x)=\hat{K}_{2} K(x)=0\;\; \mbox{and} \;\; g(x)=\hat{M}_{2} K(x)=0
$$
resolve, respectively, the equations $\hat{K}_{4} K(x) =0$\; and $\hat{M}_{4} M(x) =0$\;.
Though generally one may expect to derive in this way only a part of all solutions,
we will see that in the present case this approach allows complete solution of the problem.

It is readily checked that Eq.(\ref{4.26}) for $K(x)$  is factorized as follows:
\begin{eqnarray}
 \left [ {d^{2} \over d{x}^{2}} +{1\over 2}\, \left( {3\over x}-{7\over 1-x}\right)
  { d   \over dx} +{1\over 4}\, \left (  {p^{2}-a^{2}-10\over x}+{p^{2} -a^{2}-10\over 1-x}-{ a^{2}-6\over
   ( 1-x ) ^{2}} \right )
 \right] f ( x )=0\,,
\nonumber
\\
\label{4.33}
\end{eqnarray}

\noindent where
\begin{eqnarray}
f(x) \equiv \left [ { {d^{2}  \over d{x}^{2}}} + {1\over
2}\,\left({1\over x}-{3\over 1-x} \right) {d\over dx} + {1\over
4}\,\left ( {{p}^{2}-{a}^{2}\over x}+{ {p}^{2}-{a}^{2}\over 1-x}-{
{a}^{2}\over ( 1-x ) ^{2}}
 \right ) \right] K (x) \,.
\end{eqnarray}

\noindent Further, the equation $f(x)=0$\; is solved in terms of the hypergeometric functions. To derive this solution, we  make the
substitution
 $K (x) =x^{A}(1-x)^{B}F(x)$ with
\begin{eqnarray}
A=0\,,\;{1\over 2}\,,\qquad B=-{1\over 4}\pm{1\over
4}\,\sqrt{4a^{2}+1}\,, \quad F(x) = F(\alpha , \beta; \gamma; x)\; ,
\end{eqnarray}
\begin{eqnarray}
\alpha=A+B+{1\over 2}-{1\over 2}\,\sqrt{p^{2}+1}\,,\quad
\beta=A+B+{1\over 2}+{1\over 2}\,\sqrt{p^{2}+1}\,,\quad
\gamma={1\over 2}+2A\,.
\end{eqnarray}

\noindent Each pair $(A,B)$ of possible four such choices produces a solution of Eq.(\ref{4.26}).
However, we are interested in finite and continuous solutions at $x=0$ and $x=1$,
therefore, we take the following two pairs which produce two independent solutions of Eq.(\ref{4.26}), $K_{1}(x)$ and $K_{2}(x)$:
\begin{eqnarray}
(i) \; A={1\over 2}\,,\qquad B=-{1\over 4}+{1\over 4}\,\sqrt{4a^{2}+1} =
+ {j \over 2}   \; >  0 \,, \nonumber
\end{eqnarray}
\begin{eqnarray}
K_{1}=x^{1/2}\,(1-x)^{j/2}F\left( 1 +  j /  2  -{1\over
2}\,\sqrt{p^{2}+1}\,,  1 + j /2  +{1\over
2}\,\sqrt{p^{2}+1} \,,\;3/2 \,,\;x\right)\, ;
\end{eqnarray}
\begin{eqnarray}
(ii) \; A= 0\,,\qquad B=-{1\over 4}+{1\over 4}\,\sqrt{4a^{2}+1} = +{j
\over 2} \; >  0 \,, \quad \, \nonumber
\end{eqnarray}
\begin{eqnarray}
K_{2}=(1-x)^{j/2 } F\left(
 j/ 2  + 1/2  -{1\over 2}\,\sqrt{p^{2}+1}\,,
 j / 2 + 1 /  2 +{1\over 2}\,\sqrt{p^{2}+1}
\,,\; 1/  2 \,,\;x\right)\, .
\end{eqnarray}
It is now checked that if $K_{1}(x)$ and $K_{2}(x)$ are substituted, Eq.(\ref{eq.1}) produces
two independent functions, $M_{1}(x)$ and $M_{2}(x)$, respectively, which satisfy Eq.(\ref{4.29}) for $M(x)$.
Since, due to the hypergeometric differential equation, 
the second derivative of a hypergeometric function is expressed
in terms of this function and its first derivative,
it is readily understood by inspection of the structure of Eq.(\ref{eq.1}) that each of the functions $M_{1}(x)$ and $M_{2}(x)$
is written as a linear combination of two hypergeometric functions with rational coefficients.


In turn,  Eq. (\ref{4.29}) for $M(x)$  is factorized as
\begin{eqnarray}
 \left [ {d^{2} \over d{x}^{2}} +{1\over 2}\,
  \left( {3\over x}-{7\over 1-x}\right) { d  ( x)\over dx} +
  {1\over 4}\, \left (  {p^{2}-a^{2}-9\over x}+{p^{2} -a^{2}-9\over 1-x}-{ a^{2}-6\over  ( 1-x ) ^{2}}
\right )  \right ] g ( x )=0\,, \nonumber
\\
\label{4.42'}
\end{eqnarray}
\noindent where
\begin{eqnarray}
g(x)= \left [ { {d^{2} \over d{x}^{2}}} + {1\over
2}\,\left({1\over x}-{3\over 1-x} \right) {d\over dx} + {1\over
4}\,\left ( {{p}^{2}-{a}^{2}-1\over x}+{ {p}^{2}-{a}^{2}-1\over
1-x}-{ {a}^{2}\over ( 1-x ) ^{2}}
 \right ) \right] M (x)\,.
\end{eqnarray}
\noindent  The equation $g(x)=0$\, is also solved in terms of the hypergeometric functions. The proper substitution is
$M (x) =x^{C}(1-x)^{D}F(x)$:
\begin{eqnarray}
C=0\,,\;{1\over 2}\,,\qquad D=-{1\over 4}\pm{1\over
4}\,\sqrt{4a^{2}+1}\,, \quad F (x) = F(\alpha', \beta'; \gamma' ; x) \; ,
\\
\alpha'=C+D+ {1\over 2}- {p\over 2},\qquad \beta'=C+D+{1\over
2}+{p\over 2},\qquad \gamma'={1\over 2}+2C\,.
\end{eqnarray}

Here again there are two choices for the parameters that lead to finite solutions:
\begin{eqnarray}
(iii) \quad \quad  C= 1 /2 \,,\; D=
+ j /   2 \,,\qquad\qquad
\nonumber
\\
 M_{3}=x^{1/2}\,(1-x)^{j/2 }   F\left( 1 + j /  2   -{p\over
2} , 1 + j /  2   +{p\over 2}  \,,\; 3/2 \,,\;x\right)\, ,
\end{eqnarray}
\begin{eqnarray}
(iv) \quad  \quad \;\;C=0 \,,\;\;  D=  + j /  2 \,, \qquad \qquad
\nonumber
\\
 M_{4}=(1-x)^{j/ 2}\,
F\left( j / 2  + 1 /2   -{p\over 2}, j /2  + 1 / 2  +{p\over 2}
 \,,\; 1 / 2 ,\;x\right) \,.
\end{eqnarray}

In the similar way as was done above, it is checked that this time if $M_{3}(x)$ and $M_{4}(x)$ are substituted,
Eq.(\ref{eq.2}) produces two independent functions, respectively, $K_{3}(x)$ and $K_{4}(x)$,
which satisfy Eq.(\ref{4.26}) for $K(x)$.
It is again understood from Eq.(\ref{eq.2}) that each of the functions $K_{3}(x)$ and $K_{4}(x)$
is written as a linear combination of two hypergeometric functions with rational coefficients.

It is readily verified, e.g., by checking the Wronskian, that the functions ${K_{1},K_{2},K_{3},K_{4}}$ are independent, hence, compose a set of fundamental solutions of Eq.(\ref{4.26}) thus producing the general solution of this equation. Similarly, the functions ${M_{1},M_{2},M_{3},M_{4}}$ produce the general solution of Eq.(\ref{4.29}).

Passing from $p$ to a quantity $n$ via substitution
 $p^{2}(n)$ given as shown below, these functions are written as
\begin{eqnarray}
(i) \quad  p^{2}_{(1)} =  ( j +2 +2n)^{2} - 1 ,\quad n =0,1, 2, ... \;
, \nonumber
\\
K_{1}(x) =  \sqrt{x}\; (1-x)^{j/2}\;  F(-n,  j+2 + n ; 3 / 2;\;
x) \; , \;\; M_{1}(x)\; , \label{K1}\end{eqnarray}
\begin{eqnarray}
(ii) \quad  p^{2}_{(2)} = (j+1 + 2n )^{2} -1 , \quad n =0,1, 2, ... \;
, \nonumber
\\
 K_{2} (x) =
(1-x)^{j/2}\;  F(-n,  j+1 + n ; 1 /  2;\;  x) \; , \;\;
M_{2}(x)\;, \label{K2}\end{eqnarray}
\begin{eqnarray}
(iii) \quad  p^{2}_{(3)} = (j+2n)^{2}, \quad n =0,1, 2, ... \; ,
\nonumber
\\
 K_{3} (x), \;\; M_{3} (x) =
\sqrt{x} \; (1-x)^{j/2}\;  F(-n,  j+2 + n ; 3 / 2; \;  x) \; ,
\label{K3}\end{eqnarray}
\begin{eqnarray}
(iv) \quad  p^{2}_{(4)}  = (j +1+2n)^{2}, \quad n =0,1, 2, ... \; ,
\nonumber
\\
 K_{4} (x), \;\; M_{4} (x)  =
(1-x)^{j/2}\;  F(-n,  j+1 + n ; 1 /  2; \;  x) \; ,
\label{K4}\end{eqnarray}

\noindent where the lacking functions $M_{1}(x)$, $M_{2}(x)$  and $K_{3}(x)$, $K_{4}(x)$ are found from Eqs. (\ref{eq.1}) and (\ref{eq.2}):
\begin{eqnarray}
(i) \;  M_{1} (x) =  {( 1-x) ^{j/2} \over \,\sqrt {j ( j+1 ) }} \left(2 n (x-1) \, _2F_1\left(1-n,j+n+2;\frac{3}{2};x\right)-\right. \nonumber
\\
\left.(j x+2 n (x-1)+x-1) \, _2F_1\left(-n,j+n+2;\frac{3}{2};x\right)\right) ,
\label{MM1}
\end{eqnarray}
\begin{eqnarray}
(ii) \;  M_{2} (x) =\frac{(1-x)^{j/2}}{\sqrt{j (j+1) x}} \left(2 n (x-1) \, _2F_1\left(1-n,j+n+1;\frac{1}{2};x\right)-\right. \nonumber
\\
\left.(j x+2 n (x-1)) \, _2F_1\left(-n,j+n+1;\frac{1}{2};x\right)\right),
\label{MM2}
\end{eqnarray}
\begin{eqnarray}
(iii) \; K _{3}(x) = \frac{(1-x)^{j/2}}{\sqrt{j (j+1)}}\left(2 n (x-1) \, _2F_1\left(1-n,j+n+2;\frac{3}{2};x\right)-\right. \nonumber
\\
\left.((j+2) x+2 n (x-1)-1) \, _2F_1\left(-n,j+n+2;\frac{3}{2};x\right)\right),
\label{KK3}
\end{eqnarray}
\begin{eqnarray}
(iv) \; K _{4}(x) =\frac{(1-x)^{j/2}}{\sqrt{j (j+1)} \sqrt{x}}\left(2 n (x-1) \, _2F_1\left(1-n,j+n+1;\frac{1}{2};x\right)-\right. \nonumber
\\
\left.((j+1) x+2 n (x-1)) \, _2F_1\left(-n,j+n+1;\frac{1}{2};x\right)\right).
\label{KK4}\end{eqnarray}

For arbitrary $a$ and $p$ (i.e. if the parameters $j$ and $n$ are
mathematically considered as arbitrary complex numbers),
Eqs.(\ref{K1})-(\ref{K4}) define the general solution of the
fourth-order differential equations (\ref{4.26}) and (\ref{4.29}).

Since the derived general solution involves combinations of
hypergeometric functions with rational coefficients, the discrete
energy spectrum can be obtained by terminating the involved
hypergeometric series. Indeed, this procedure necessarily results
in quasi-polynomial solutions, which warrant discrete energy
levels owing to the above-chosen particular asymptotes for the
corresponding pre-factors for functions $K_{i},M_{i}$. The
condition for termination of a hypergeometric series is that an
upper parameter of it is a negative integer. This is achieved if
$n =0,1, 2, ... \;$. Correspondingly, Eqs.(\ref{K1})--(\ref{K4})
then define four sets of quasi-polynomial wave functions
describing the discrete energy states of a Dirac-K\"{a}hler
particle.
The derived solutions reveal significant differences
as compared with the result for the ordinary Dirac particle in spherical space.

The explicit form of the Dirac equation in the space model
discussed here is
\begin{eqnarray}
\left (  i \gamma^{0} {\partial \over \partial t } \;  + \; i
\gamma^{3} {\partial \over \partial r } \; + \; {1 \over \sin r }
\Sigma_{\theta \phi}\; -\;  m   \right )  \Psi = 0 \
\label{D-1}
\end{eqnarray}

\noindent and  the general structure of the wave function
corresponding to the diagonalization of operators  $i\partial_{t},
\vec{J}^{2}, J_{3}$ is
\begin{eqnarray}
\tilde{\Psi} = e^{-i \epsilon t } \left | \begin{array}{r}
f_{1}(r) \; D_{-1/2} \\
f_{2}(r) \; D_{+1/2} \\
f_{3}(r) \; D_{-1/2} \\
f_{4}(r) \; D_{+1/2}
\end{array} \right | ,
\label{D-2}
\end{eqnarray}

\noindent where the Wigner functions \cite{VMX} are denoted as
$D_{\sigma}= D^{J}_{-m,\sigma}(\phi, \theta,0)$;   $J$ adopts half-integer values: $J =1/2, 3/2,
...\;$.

Omitting the details (see the discussion in \cite{book-1}), the
result for the  energy spectrum of the Dirac particle is given as
\begin{eqnarray}
p^{2} = \epsilon^{2} - m^{2} =  (n +  J +1)^{2}  , \label{D-3}
\end{eqnarray}
or in usual units
\begin{eqnarray}
E^{2} - m^{2} c^{4} =  {\hbar^{2} c^{2} \over  \rho^{2}} \; (n + J
+1  )^{2}  \; , \label{D-3'}
\end{eqnarray}

\noindent where $\rho$ stands for the curvature radius.

 Examining  the derived spectrum (\ref{K1})-(\ref{K4})
for the Dirac--K\"{a}hler particle, we note that all the four energy series differ from the
energy spectrum  (\ref{D-3}) for the Dirac particle. Further, we note that
$p^{2}_{(2)} \to p^{2}_{(1)}$ and $p^{2}_{(4)} \to p^{2}_{(3)}$ if
$j$ is shifted by unity: $j \to j+1$ . However, since the
corresponding wave functions remain distinct, and thus present
different states, we see that the energy spectrum of a
Dirac-K\"{a}hler particle in a Riemann spherical space consists of
two paralleled series each of which is twofold degenerate.

\section{Discussion}

Thus, we have shown that  the energy spectrum of the Dirac-K\"{a}hler
particle in spherical space is much more complicated compared with
the spectrum for the Dirac particle. The difference is of both
structural and quantitative character. The energy spectrum of a
Dirac-K\"{a}hler particle consists of two paralleled series each
of which is twofold degenerate.
Besides, none of the separate energy level series coincides with
that one for the Dirac particle.

We conclude with a brief discussion of some additional points concerning the relationship
between the Dirac--K\"{a}her field and a set of four Dirac fields
in the case of flat Minkowski space. In order to explicitly
connect the  boson solutions of the Dirac--K\"{a}hler field
with spherical solutions of the (four) Dirac equations, one should
perform a special transformation $U(x) \; \rightarrow \; V(x)$
so chosen that in the new representation the Dirac--K\"{a}hler equation
is splitted into four separated Dirac-type equations (see \cite{Ovsiyuk-Kisel-Redkov} for more details).
Then, it is possible to decompose the four columns of the $(4 \times 4)$-matrix  $V(x)$,
related with the Dirac--K\"{a}hler  equation,
in terms of the solutions of four Dirac-type equations.
The mentioned transformation has the form
\begin{eqnarray}
V(x)  = ( I \otimes  S(x))\; U(x) \; , \qquad  S(x)  = \left |
\begin{array}{cc}  U(x) & 0 \\ 0 & U(x)
\end{array} \right |   ,
\nonumber
\\
U(x) =   \left |  \begin{array}{cc}
\cos {\theta \over 2} \; e^{-i\phi /2} & \sin  {\theta \over 2} \; e^{-i\phi /2} \\[2mm]
-\sin  {\theta \over 2} \; e^{+i\phi /2} & \cos {\theta \over 2}\;
e^{+i\phi /2}
\end{array}  \right |      . \qquad\qquad
\label{a.1}
\end{eqnarray}

\noindent The spherical bispinor connection  $\Gamma _{\alpha }$:
$$
\Gamma _{t} = 0 \; , \;\; \Gamma _{r} = 0 \; , \;\; \Gamma
_{\theta } = j^{12} \; ,  \;\; \Gamma _{\phi } =  \sin \theta \;
j^{32}  +  \cos \theta \j^{12} \; ,
$$

\noindent involved in   the Dirac--K\"{a}hler connection
$ B_{\alpha} (x) = \Gamma_{\alpha}(x) \otimes I + I
\otimes \Gamma_{\alpha}(x) $ (see Eq.(\ref{1.2}))
in the  $V(x)$-representation is written as
\begin{eqnarray}
 \{ \; [\; i \gamma ^{\alpha }(x) \; ( \partial_{\alpha } \; + \;
\Gamma _{\alpha }(x))  \; - \; m \;  ] \; V(x) + \qquad \qquad
\nonumber
\\
 + i\; \gamma ^{\alpha }(x) \; V(x) \; [\; S(x) \; \Gamma
_{\alpha}(x) \; S^{-1}(x) \; + \; S(x) \; \partial _{\alpha}  \;
S^{-1}(x) \; ]\;  \}  = 0\; .
\nonumber
\end{eqnarray}

\noindent The chosen form  (\ref{a.1})  of $S(x)$ leads to vanishing the term
$$
 S(x)\; \Gamma _{\alpha }(x)\; S^{-1}(x) \; + \;
S(x)\; \partial_{\alpha }\;  S^{-1}(x) = 0 \; .
$$

\noindent
With this, we finally arrive at four disconnected Dirac equations for the columns of the matrix $V(x)$:
\begin{eqnarray}
\left [ \; i \gamma ^{\alpha }(x)\; (\partial _{\alpha }\; + \;
\Gamma _{\alpha }(x)) \; -  \; m \; \right ]\;  V(x) = 0\; .
\label{a.3}
\end{eqnarray}

Considering now the above spherical Dirac--K\"{a}hler solution, we thus should transform
the solution from the $U$-form to a corresponding $V$-form, and second,
should expand the four columns of the matrix $V$ in terms of the Dirac spherical waves.
The calculations performed in \cite{Ovsiyuk-Kisel-Redkov} lead to simple linear expansions
of the four columns of the new representative of the Dirac--K\"{a}hler field
$V(x)$ in terms of spherical fermion solutions $\Psi_{i}(x)$  of
the four ordinary Dirac equations.
It should be emphasized, however, that this procedure for construction
of such expansions is applicable only for the case of flat Minkowski
space-time; it cannot be extended to any space-time model with curvature.

Moreover, even in the case when such expansions exist,
this fact cannot be interpreted
as the equivalence of the Dirac--K\"{a}hler field
and a system of four Dirac fermions.
An essential objection against such an equivalence is that the mentioned
transformation $(S(x) \otimes I)$ does not belong to the
group of tetrad local gauge transformations for Dirac--K\"{a}hler field,
which is a 2-rank bispinor under the Lorentz group.
Therefore, the linear expansions between boson and fermion functions are not
gauge invariant under the group of local tetrad rotations.

Finally, it should be also emphasized that the Wigner $D$-functions,
related to integer and half-integer values of $j(J)$,
involved, respectively, in the Dirac-K\"{a}hler wave functions $U(x)$ and the Dirac functions $\Psi (x)$,
exhibit completely different boundary properties in angular variables $(\theta, \phi)$
(see the details in \cite{VMX}).

\section{Acknowledgements}
This work was supported by the Fund for Basic Researches of Belarus
(Grant No. F14ARM-021) and by the Armenian State Committee of Science (Grant No. 13RB-052),
within the cooperation framework between Belarus and Armenia.
The research has been partially conducted within the scope of the International
Associated Laboratory (CNRS-France \& SCS-Armenia) IRMAS.
The work by A. Ishkhanyan has received funding
from the European Union Seventh Framework Programme, grant No. 295025 - IPERA).

\newpage

Artur Ishkhanyan

Institute for Physical Research, Armenia

E-mail: aishkhanyan@gmail.com

\bigskip

Olivia Florea


Transilvania University of Brasov,


E-mails: olivia.florea @unitbv.ro

\bigskip

Elena Ovsiyuk

Mozyr State Pedagogical University, Belarus.

E-mail: e.ovsiyuk@mail.ru

\bigskip

Viktor Red'kov

Laboratory of Theoretical Physics, Institute of Physics,

National Academy of Sciences of Belarus.

E-mail: v.redkov@dragon.bas-net.by

\end{document}